# A Paradox Related to Mechanical and Electrical Energy Conversion

Songyan Li

Fulton School of Engineering, Arizona State University, Tempe

**Abstract**

A model is constructed and a paradox concerning the proper direction and magnitude of the external force to maintain the equilibrium state of a parallel plate capacitor system is raised. By a thorough study of the model using different methods, the paradox is eliminated.

## 1. Introduction

Several interesting paradoxes related to parallel capacitors had been raised[1][2][3][4]. No matter which method which we use to analysis a physical system, as long as we are on the right tracks, the equations describing the system should be the same. Because the nature of the physical system should not change while we are changing our ways of studying the system. However, sometimes even for a simple system, the manner of the motion of the system does seems to change depending on the point of view we choose to examining the system. When this happens, it encourages us to do a thorough study of it. Historically such events were happening through out the development of physics, such as Maxwell's Demon and EPR paradox[5]. The process of eliminating a paradox often deepens our understanding about the nature. In this article, a paradox about mechanical energy and electrical energy conversion is raised.

## 2. The Model and the Paradox

Suppose we have a parallel plate capacitor in a vacuum environment, as shown in Figure 2.1. We restrict the two plates that it can only move horizontally and their direction of displacement can only lie in the surface of the page. We also neglected the edge effect. The plate on the right side is movable while the plate on the left side is fixed. Neglect all frictional force. We apply a voltage source with the potential difference across the source U fixed. Those two plates have same area $A$. The left capacitor is located at the origin, and the right capacitor is located at point x. The object is to find out the direction and magnitude of the external force that can make the system stationary.

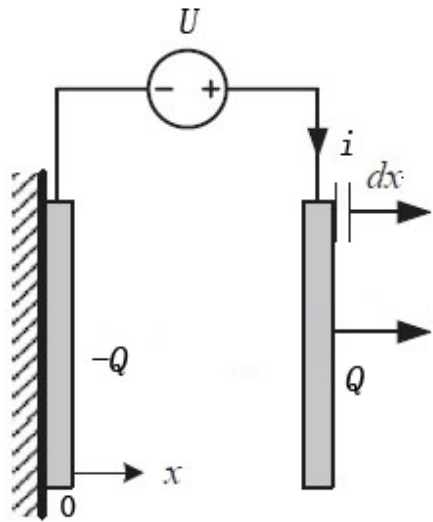

Figure 2.1

**2.1. Idea 1**

We now concern about the energy stored in the system, that is, the energy stored in the electric field. Notice that the capacitance of the capacitor is[6]

$$C = \frac{\varepsilon_0 A}{x} \qquad (2.1)$$

According to the well-known formula about energy stored in the capacitor[6]

$$E_e = \frac{CU^2}{2} \qquad (2.2)$$

$$E_e = \frac{\varepsilon_0 A U^2}{2x} \qquad (2.3)$$

It will be decreased when there is a differential displacement dx increasing the distance between the two plates. If the system is spontaneously moving into a state that has lower energy, the system will move to some state that has smaller C, hence smaller energy stored in the field, while U is fixed. That is, the two plates are trying to separate from each other. The tendency of the motion of the plate on the right side is moving away from another plate. Thus we need to apply an external force pointing to the left in order to maintain an equilibrium state of the system. This is a common idea held by several scholars in some textbooks[7]. The writer of this article found this is also a common idea held by many graduate students in his university.

**2.2. Idea 2**

We start from the electric force acting on the right plate. According to the potential difference of the two plates, the direction of electric field, E, is pointing from the right to the left. Thus, the right plate, with positive charge on it, is subjected to a electric force pointing to the left. So we need to apply an external force pointing to the right to maintain the equilibrium state.

Now, the paradox is raised. Idea 1 tells us the force should point to the left and Idea 2 tells us the force should point to the right. What is happening here?

## 3. What is the trick?

We don't apply any external force on the movable plate and study it's motion. As long as the right plate is movable, we can never assume the kinetic energy of the plate is unchanged during the change of distance between the two plates. Therefore, the kinetic energy should be considered as a part of the total energy of the system. By definition[8], the kinetic energy of the movable plate is

$$E_k = \frac{m\dot{x}^2}{2} \tag{3.1}$$

The total energy of the system is then the kinetic energy of the movable plate plus the energy stored in the field.

$$E_{system} = E_k + E_e = \frac{m\dot{x}^2}{2} + \frac{\varepsilon_0 A U^2}{2x} \tag{3.2}$$

where $m$ is assumed to be the mass of the movable plate. Notice that the voltage source can do work during the process because moving the plate changes the capacitance of the capacitor hence the changing charge of the plate generate current that flow through the voltage source. By definition, the work done by the source is

$$dW_{source} = Uidt \tag{3.3}$$

where the current, as shown in the Figure 2.1, is

$$i = \frac{dq}{dt} \tag{3.4}$$

So the work done by the source is then

$$dW_{source} = Udq = Ud(CU) = U^2 dC + UCdU \tag{3.5}$$

We assumed that the voltage across the source is fixed, so

$$dU = 0 \tag{3.6}$$

and

$$dW_{source} = U^2 dC = -\frac{\varepsilon_0 A U^2}{x^2} dx \tag{3.7}$$

According to the relationship between the work done to the system and the differential change of system energy, we have

$$dW_{source} = dE_{system} \tag{3.8}$$

which, by (3.7) and (3.2), leads to

$$-\frac{\varepsilon_0 A U^2}{x^2} dx = d\left(\frac{m\dot{x}^2}{2} + \frac{\varepsilon_0 A U^2}{2x}\right) \qquad (3.9)$$

We expand the term on the right side of (3.9)

$$-\frac{\varepsilon_0 A U^2}{x^2} dx = m\dot{x}d\dot{x} - \frac{\varepsilon_0 A U^2}{2x^2} dx \qquad (3.10)$$

The term

$$m\dot{x}d\dot{x}$$

can be expand to

$$m\dot{x}d\dot{x} = m\frac{dx}{dt}d\dot{x} \qquad (3.11)$$

Substitute (3.11) into (3.10), we get

$$m\frac{d\dot{x}}{dt} = -\frac{\varepsilon_0 A U^2}{2x^2} \qquad (3.12)$$

What does this mean? Newton's Law tells us the net force acting on the movable plate is

$$F_{net} = -\frac{\varepsilon_0 A U^2}{2x^2} \qquad (3.13)$$

where the minus sign means the net force acting on the plate is pointing to the left, which is NOT the result provided by Idea 1. So if the movable plate is stationary at the beginning, it will move toward the fixed plate.

In this process, the energy of the system is increasing. The movable plate is starting to move toward the other plate, increasing the kinetic energy and the energy stored in the field is also increasing by (2.2). The reason why the system is not spontaneously moving to a lower energy state, is that the system is actually not a closed system, because there is a voltage source and the non-electro static force of the voltage source do work to the system. It can possibly increase the energy of the whole system we are concerning. The work done by the source will partially become the kinetic energy of the movable plate while partially become the energy stored in the field of the capacitor.

An analogue of a pure mechanical model may be illustrated to make it easier to understand. Suppose that we have an iron ball in the gravitational field near the surface of the earth. If we apply a upward force to the ball that is larger than the gravitational force of the ball, the ball will move upward and accelerate. Hence the total energy, which is the sum of potential energy and kinetic energy of the ball, will be larger than the original state. Where is this additional energy come from? The force acting on the ball does the work.

**4. Further Analysis of the Model Focusing on the External Force**

If we want to apply an external force F to make the movable plate stationary, what force should we apply? We again use the relationship between the work done to the system and the change of the energy of the system. But this time the external force do the work

$$dW_{ex} = Fdx \tag{4.1}$$

$$dW_{ex} + dW_{source} = dE_{system} \tag{4.2}$$

which leads to

$$Fdx - \frac{\varepsilon_0 AU^2}{x^2}dx = m\dot{x}d\dot{x} - \frac{\varepsilon_0 AU^2}{2x^2}dx \tag{4.3}$$

We use (3.11) again, and get the equation

$$m\frac{d\dot{x}}{dt} = F - \frac{\varepsilon_0 AU^2}{2x^2} \tag{4.4}$$

If the movable plate is rest at the beginning, the condition to make it remain it's original state is that the net force acting on it

$$F_{net} = F - \frac{\varepsilon_0 AU^2}{2x^2} \tag{4.5}$$

must be zero. The external force acting on the movable plate is obtained.

$$F = \frac{\varepsilon_0 AU^2}{2x^2} \tag{4.6}$$

The equation above shows that $F$ must be positive and it means the external force must point to the right.

**5. The Comparison to a Direct Analysis of the Electric Force Acting on the Plate**

From electromagnetism we have the formula of the electric field generated by uniformly distributed negative charge on the left plate in the space that is between the two plates[6]

$$E = \frac{-\sigma}{2\varepsilon_0} \tag{5.1}$$

where $\sigma$ is the surface charge density. The electric field should point to the left because the right plane has a higher potential. The minus sign represents that the direction of this field is pointing to the left.

$$\sigma = \frac{Q}{a} \tag{5.2}$$

where in our case

$$Q = CU \tag{5.3}$$

Every charge on the right plate is in the electric field generated by the charge on the left plate. Therefore the electric force acting on the right plate is just

$$F_e = EQ \tag{5.4}$$

By (5.1), (5.2), (5.3), we have

$$F_e = EQ = -\frac{\sigma}{2\varepsilon_0}CU = -\frac{1}{2\varepsilon_0 A}C^2U^2 = -\frac{\varepsilon_0 AU^2}{2x^2} \tag{5.5}$$

This is just the same force that (3.13) represents, which again examined our analysis.

## 6. How About a Changing U?

We considered the model with a fixed voltage source. Again, we don't apply the external force. What will happen if the voltage U of the source is changing with time? The answer is at any given time t, the net force acting on the movable plate is still

$$F_{net}(t) = -\frac{\varepsilon_0 A[U(t)]^2}{2x^2} \tag{6.1}$$

where the minus sign means the force is pointing toward left. This can be justified by the following process. The work done by the source during the process is

$$dW_{source} = Udq = Ud(CU) = U^2 dC + UCdU \tag{6.2}$$

$$dW_{source} = dE_{system} \tag{6.3}$$

$$U^2 dC + UCdU = d\left(\frac{m\dot{x}^2}{2}\right) + d\left(\frac{CU^2}{2}\right) \tag{6.4}$$

$$U^2 dC + UCdU = m\dot{x}d\dot{x} + \frac{U^2}{2}dC + CUdU \tag{6.5}$$

Two $UCdU$ terms cancelled and by (2.1) it leads to

$$F_{net} = m\frac{d\dot{x}}{dt} = -\frac{\varepsilon_0 AU^2}{2x^2} \tag{6.6}$$

again. Still the net force acting on the right plate is pointing to the left plate. And obviously an analysis of the electric force similar to section 5 would yield the same result.

## 7. Conclusion

It has been clear what is the problem of Idea 1. Because the system is not closed, we have to consider the work done by the source and also be cautious to take the kinetic energy into account. Therefore the paradox is eliminated. Problems often seems to be easier than they are after we find out the tricks. But from the process we study this paradox, we deepen our understanding of classical mechanics and electromagnetism.

# References

[1] A. Singal, arXiv:1309.5034, [gen-ph]

[2] V. Pankovic, arXiv:0912.0648, [gen-ph]

[3] K. T. McDonald, arXiv:physics/0312031, [class-ph]

[4] G. D'Abramo, arXiv:0912.4818, [gen-ph]

[5] D. J. Griffith, Introduction to Quantum Mechanics (2nd Edition), Pearson Prentice Hall, Apr 2004, p421-423

[6] D. J. Griffith, Introduction to Electrodynamics (3rd Edition), Addison Wesley, Jan 1999, p193-195, p73-74 ,p102-103

[7] Karady, G. and Holbert, K. Electrical Energy Conversion and Transport. Wiley-Blackwell. Jul 2013, p344-345

[8] Goldstein, H. Poole, C and Safko, J. Classical Mechanics. Addison-Wesley. Jun 2001, p3-4